\documentclass[12pt,letterpaper]{article}

\usepackage[]{graphicx}\usepackage[]{color}
\makeatletter
\def\maxwidth{ %
  \ifdim\Gin@nat@width>\linewidth
    \linewidth
  \else
    \Gin@nat@width
  \fi
}
\makeatother

\definecolor{fgcolor}{rgb}{0.345, 0.345, 0.345}
\newcommand{\hlnum}[1]{\textcolor[rgb]{0.686,0.059,0.569}{#1}}%
\newcommand{\hlstr}[1]{\textcolor[rgb]{0.192,0.494,0.8}{#1}}%
\newcommand{\hlopt}[1]{\textcolor[rgb]{0,0,0}{#1}}%
\newcommand{\hlstd}[1]{\textcolor[rgb]{0.345,0.345,0.345}{#1}}%
\newcommand{\hlkwc}[1]{\textcolor[rgb]{0.333,0.667,0.333}{#1}}%
\newcommand{\hlkwd}[1]{\textcolor[rgb]{0.737,0.353,0.396}{\textbf{#1}}}%

\usepackage{framed}
\makeatletter
\newenvironment{kframe}{%
 \def\at@end@of@kframe{}%
 \ifinner\ifhmode%
  \def\at@end@of@kframe{\end{minipage}}%
  \begin{minipage}{\columnwidth}%
 \fi\fi%
 \def\FrameCommand##1{\hskip\@totalleftmargin \hskip-\fboxsep
 \colorbox{shadecolor}{##1}\hskip-\fboxsep
     \hskip-\linewidth \hskip-\@totalleftmargin \hskip\columnwidth}%
 \MakeFramed {\advance\hsize-\width
   \@totalleftmargin\z@ \linewidth\hsize
   \@setminipage}}%
 {\par\unskip\endMakeFramed%
 \at@end@of@kframe}
\makeatother

\definecolor{shadecolor}{rgb}{.97, .97, .97}
\definecolor{messagecolor}{rgb}{0, 0, 0}
\definecolor{warningcolor}{rgb}{1, 0, 1}
\definecolor{errorcolor}{rgb}{1, 0, 0}
\newenvironment{knitrout}{}{} 

\usepackage{alltt}

\usepackage{amssymb,amsmath}
\usepackage[export]{adjustbox}
\usepackage{stix}

\providecommand{\tightlist}{%
  \setlength{\itemsep}{0pt}\setlength{\parskip}{0pt}}

\usepackage{textcomp}

\makeatletter

\newcommand\midtilde@raisedtilde[1][.5]{\raisebox{#1ex}{\texttildelow}}
\def\midtilde@normaltilde{\texttildelow}

\newcommand\midtilde%
{%
  \expandafter\in@\expandafter{\f@family}%
    {cmr,cmss,cmtt,cmm,cmsy,cmx,%
    lmr,lmss,lmtt,lmm,lmsy,lmx,%
    pxr,pxss,pxm,pxsy,pxx,%
    txr,txss,txm,txsy,txx}%
  \ifin@%
    \midtilde@raisedtilde%
  \else%
    \expandafter\in@\expandafter{\f@family}%
    {pxtt,txtt}%
    \ifin@%
      \midtilde@raisedtilde[.35]%
    \else%
      \midtilde@normaltilde%
    \fi%
  \fi%
}

\usepackage{enumitem}
\usepackage[round]{natbib}
\usepackage{color}
\usepackage{fancyvrb}

\DefineVerbatimEnvironment{Highlighting}{Verbatim}{commandchars=\\\{\}}
\usepackage{framed}
\definecolor{shadecolor}{RGB}{248,248,248}

\usepackage{xspace}
\newcommand{\mytilde}{\lower.80ex\hbox{\char`\~}}

\usepackage{graphicx}
\usepackage{booktabs}
\usepackage{hyperref}
\title{Symbolic Formulae for Linear Mixed Models\thanks{Supported by R Consortium}}
\author{Emi Tanaka\textsuperscript{1,2,*} \and Francis K. C. Hui\textsuperscript{3}}
\date{%
$^{1}$ School of Mathematics and Statistics, The University of Sydney, NSW, Australia, 2006
\\
$^{2}$ Department of Econometrics and Business Statistics, Monash University, VIC, Australia, 3800
\\
$^{3}$ Research School of Finance, Actuarial Studies \& Statistics, Australian National University, ACT, Australia, 2601
\\[7mm]
$^{*}$ emi.tanaka@monash.edu
}

\begin{document}

\maketitle              

\begin{abstract}
A statistical model is a mathematical representation of an often simplified or idealised data-generating process. In this paper, we focus on a particular type of statistical model, called linear mixed models (LMMs), that is widely used in many disciplines e.g.~agriculture, ecology, econometrics, psychology. Mixed models, also commonly known as multi-level, nested, hierarchical or panel data models, incorporate a combination of fixed and random effects, with LMMs being a special case. The inclusion of random effects in particular gives LMMs considerable flexibility in accounting for many types of complex correlated structures often found in data. This flexibility, however, has given rise to a number of ways by which an end-user can specify the precise form of the LMM that they wish to fit in statistical software. In this paper, we review the software design for specification of the LMM (and its special case, the linear model), focusing in particular on the use of high-level symbolic model formulae and two popular but contrasting \texttt{R}-packages in \texttt{lme4} and \texttt{asreml}.
\end{abstract}

\section{Introduction}\label{introduction}

Statistical models are mathematical formulation of often simplified real world phenomena, the use of which is ubiquitous in many data analyses. These models are fitted or trained computationally, often with practitioners using some readily available application software package. In practice, statistical models in its mathematical (or descriptive) representation would require translation to the right input argument to fit using an application software. The design of these input arguments (called application programming interface, API) can help ease the friction in fitting the user's desired model and allow focus on important tasks, e.g.~interpreting or using the fitted model for purposes downstream.

While there are an abundance of application software for fitting a variety of statistical models, the API is often inconsistent and restrictive in some fashion. For example, in linear models, the intercept may or may not be included by default; and the random error typically assumed to be identical and independently distributed (i.i.d) with no option to modify these assumptions straightforwardly. Some efforts have been made in this front such as by the \texttt{parsnip} package \citep{Kuhn2018} in the \texttt{R} language \citep{R2018} to implement a tidy unified interface to many predictive modelling functions (e.g.~random forest, logistic regression, linear regression etc) and the \texttt{scikit-learn} library \citep{scikit-learn} for machine learning in the \texttt{Python} language \citep{van1995python} that provides consistent API across its modules \citep{sklearn_api}. There is, however, little effort on consistency or discussion for the software specification of many other types of statistical models, including the class of linear mixed models (LMMs), which is the focus of this article.

LMMs (a special case of mixed models in general, which are also sometimes referred to as hierarchical, panel data, nested or multi-level models) are widely used across many disciplines (e.g.~ecology, psychology, agriculture, finance etc) due to their flexibility to model complex, correlated structures in the data. This flexibility is primarily achieved via the inclusion of random effects and their corresponding covariance structures. It is this flexibility, however, that results in major differences in model specification between software for LMMs. In \texttt{R}, arguably the most popular general purpose package to fit LMMs is \texttt{lme4} \citep{Bates2015} -- total downloads from RStudio Comprehensive R Archive Network (CRAN) mirror from \texttt{cranlogs} \citep{cranlog} indicate there were over two million downloads for \texttt{lme4} in the whole of 2018, while other popular mixed model packages e.g. \texttt{nlme}, \texttt{rstan}, and \texttt{brms} \citep{nlme, rstan, brmsjss} in the same year have less than half a million downloads, albeit \texttt{rstan} and \texttt{brms} are younger packages. Another general purpose LMM package is \texttt{asreml} \citep{Butler2009}, which wraps the proprietary software ASreml \citep{Gilmour2009} into the \texttt{R} framework. As this package is not available on CRAN, there are no comparable download logs, although, citations of its technical document indicates popular usage particularly in the agricultural sciences. In this paper, we discuss only \texttt{lme4} and \texttt{asreml} due to their active maintenance, maturity and contrasting approaches to LMM specification.

The functions to fit LMM in \texttt{lme4} and \texttt{asreml} are \texttt{lmer} and \texttt{asreml}, respectively. Both of these functions employ high-level symbolic formulae as part of their API to specify the model. In brief, symbolic model formulae define the structural component of a statistical model in an easier and often more accessible terms for practitioners. The earlier instance of symbolic formulae for linear models was applied in Genstat \citep{genstat} and GLIM \citep{GLIM}, with a detailed description by \citet{Wilkinson1973}. Later on, \citet{Chambers1993} describe the symbolic model formulae implementation for linear models in the \texttt{S} language, which remains much the same in the \texttt{R} language. While the symbolic formula of linear models generally have a consistent representation and evaluation rule as implemented in \texttt{stats::formula}, this is not the case for LMMs (and mixed models more generally) -- the inconsistency of symbolic formulae arises primarily in the representation of the random effects, with the additional need to specify the covariance structure of the random effects as well as structure of the associated model matrix that governs how the random effects are mapped to (groups of) the observational units.

In Section \ref{lm}, we briefly describe the symbolic formula in linear models. We then describe the symbolic model formula employed in the LMM functions \texttt{lmer} and \texttt{asreml} in Section \ref{lmm}. We follow by illustrating a number of statistical models motivated by the analysis of publicly available agricultural datasets, with corresponding API for \texttt{lmer} and \texttt{asreml} in Section~\ref{examples}.  We limit the discussion of symbolic model formulae to mostly those implemented in \texttt{R}, however, it is important to note that the conceptual framework is not limited to this language. We conclude with a discussion and some recommendations for future research in Section~\ref{discussion}.

\hypertarget{lm}{%
	\section{Symbolic Formulae for Linear Models}\label{lm}}

A special case of LMMs is linear models, which comprises of only fixed effects and a single random term (i.e.~the error or noise), given in a matrix notation as
\begin{equation}
\boldsymbol{y} = \mathbf{X}\boldsymbol{\beta} + \boldsymbol{e},\label{eq:lm}
\end{equation}
where \(\boldsymbol{y}\) is a \(n\)-vector of responses, \(\mathbf{X}\) is the \(n\times p\) design matrix with an associated \(p\)-vector of fixed effects coefficients \(\boldsymbol{\beta}\), and \(\boldsymbol{e}\) is the \(n\)-vector of random errors. Typically we assume \(\boldsymbol{e} \sim N(\boldsymbol{0}, \sigma^2\mathbf{I}_n)\).

The software specification of linear model is largely divided into two approaches: (1) input of arrays for the response \(\boldsymbol{y}\) and design matrix for fixed effects \(\mathbf{X}\), and (2) input of a symbolic model formula along with a data frame that define the variables in the formula. The input of the data frame may be optional if the variables are defined in the parental environment, although such approach is not recommended due to larger potential for error (e.g.~one variable is sorted while others are not).

Symbolic model formulae have been heavily used to specify linear models in \texttt{R} since its first public release in 1993, inheriting most of its characteristic from \texttt{S}. In \texttt{R}, formulae have a special class \texttt{formula}, and can be used for other purposes other than model specification, such as \texttt{case\_when} function in \texttt{dplyr} \texttt{R}-package \citep{dplyr}, which uses the left hand side (LHS) to denote cases to substitute with given value on the right hand side (RHS) - these type of use is not within the scope of this paper. The history of the \texttt{formula} class in \texttt{R} (and \texttt{S}) is considerably longer than other popular languages, e.g.~the \texttt{patsy} \texttt{Python} library \citep{patsy}, which imitates \texttt{R}'s formula, was introduced in 2011 and used in \texttt{Statsmodels} library \citep{seabold2010statsmodels} to fit a statistical model.

Symbolic model formulae makes use of the variable names defined in the environment (usually through the data frame) for specifying the precise model formulation. With linear models, the LHS indicate the response; the RHS resembles its mathematical form; and the LHS and RHS are separated by \texttt{\midtilde} which can be read as ``modelled by''. For example, the symbolic model formula \texttt{y \midtilde\ 1 + x} can be thought of as the vector \texttt{y} is modelled by a linear predictor consisting of an overall intercept term and the variable \texttt{x}.

\begin{figure}
	\includegraphics[width=0.9\linewidth,fbox]{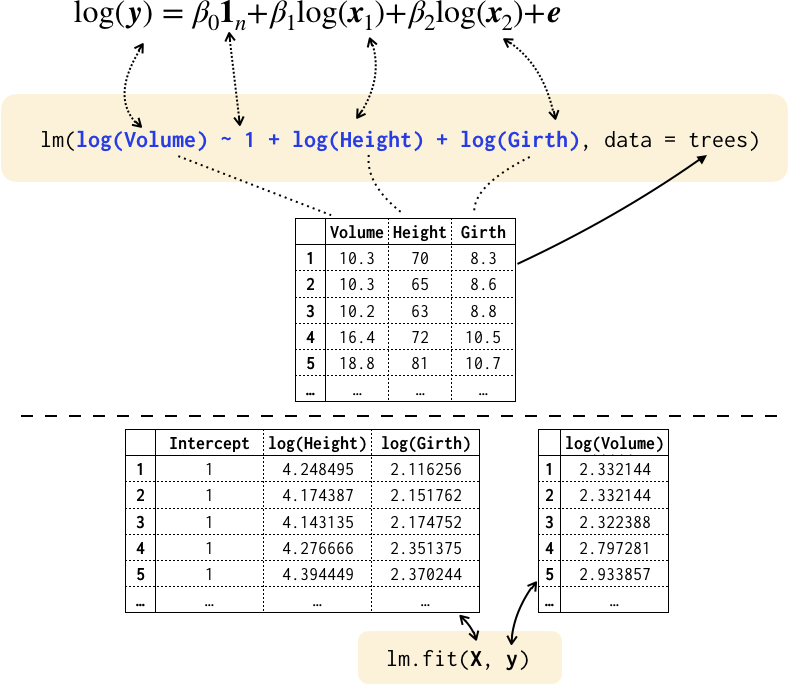} \caption{There are two main approaches to fitting a linear model illustrated above with the fit of a linear model to the \texttt{trees} dataset: (1) the top half uses the \texttt{lm} function  with the input argument as a symbolic model formulae (in blue); (2) the bottom half uses the \texttt{lm.fit} function which requires input of design matrix and the response. The latter approach is not commonly used in \texttt{R}, however, it is the common approach in other languages; see Section \ref{trees} about the data and the model.}\label{fig:symbolic-lm}
\end{figure}

When the variables are numerical then the connection between the formula to its regression equation is obvious -- the LHS is \(\boldsymbol{y}\), while the RHS corresponds to the columns of the design matrix \(\mathbf{X}\) in the linear model \eqref{eq:lm}. One advantage of this symbolic model formula approach is that any transformation to the variable can be parsed in the model formula and may be used later in the pipeline (e.g.~prediction in its original scale). This contrasts to when the input arguments are the design matrix and the corresponding response vector -- there is now an additional step required by the user to transform the data before model fitting and subsequently afterwards for extrapolation. Such manual transformation also likely results in manual back-transformation later in the analysis pipeline for interpretation reasons. This no doubt creates extra layer of friction for the practitioner in their day-to-day analysis. Figure \ref{fig:symbolic-lm} illustrates this connection using the \texttt{trees} dataset.

The specification of the intercept by \texttt{1} in the formula, as done in Figure \ref{fig:symbolic-lm}, is unnecessary in \texttt{R} since this is included by default. In turn, the removal of the intercept can be done by including \texttt{-1} or \texttt{+0} on the RHS. In this paper, the intercept is explicitly included as the resemblance to its model equation form is lost without it. While the omission of \texttt{1} is long ingrained within \texttt{R}, we recommend to explicitly include \texttt{1} and do not recommend designing software to require explicit specification to remove intercept as currently required in \texttt{R}; see Section \ref{intercept} for further discussion on this.

\begin{figure}
	\includegraphics[width=0.9\linewidth,fbox]{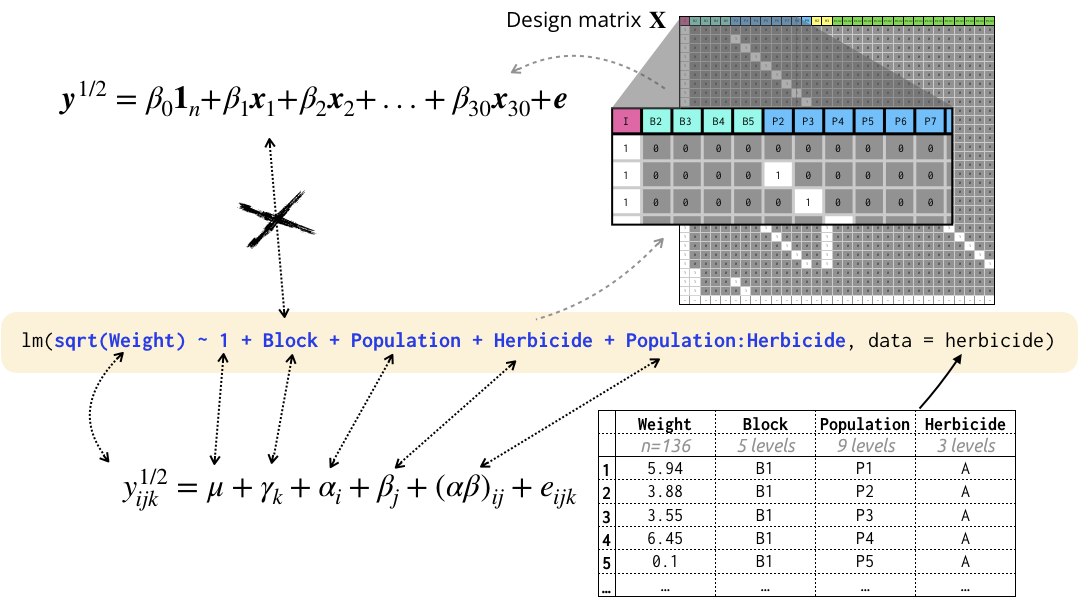} \caption{In the presence of categorical variables, the resemblance of the symbolic model formulae to its regression model form is not immediately obvious. In this case, categorical variables are transformed to a set of dummy variables with constraint applied for parameter identifiability. As such, a single categorical variable span a number of columns in the design matrix. On the other hand, if the model equation is written using the ANOVA model specification (with index notation), then the categorical variables have an immediate connection to the fixed effects in the model; see \ref{grass} for more information about the data and the model.}\label{fig:symbolic-lm-factor}
\end{figure}

Categorical or factor variables are typically converted to a set of dummy variables consisting of 0s and 1s indicating whether the corresponding observation belongs to the respective level. For parameter identifiability, a  constraint needs to be applied, e.g.~the treatment constraint will estimate effects in comparison with a particular reference group (the default behaviour in \texttt{R}). Note that in the presence of categorical variables, the direct mapping of the symbolic formula to the regression equation is lost. However, the mapping is clear in converting the model equation to the so-called Analysis of Variance (ANOVA) model specification as illustrated in Figure \ref{fig:symbolic-lm-factor}, which represents the fit of a two-way factorial ANOVA model to the \texttt{herbicide} data.

Interaction effects are specified easily with symbolic model formula by use of the \texttt{:} operator as seen in Figure \ref{fig:symbolic-lm-factor}. More specifically, the formula in Figure \ref{fig:symbolic-lm-factor} can also be written more compactly as \texttt{sqrt(Weight)} \texttt{\midtilde\ }\texttt{1\ +\ Block\ +\ Population\ *\ Herbicide} where the \texttt{*} operator is a shorthand for including both main effects and the interaction effects. Further shorthand exists for higher order interactions, e.g. \texttt{y \midtilde\ }\texttt{1\ +\ (x1\ +\ x2\ +\ x3)\^{}3} is equivalent to
\texttt{y \midtilde\ }\texttt{1\ +\ x1\ +\ x2\ +\ x3\ +\ x1:x2\ +\ x1:x3\ +\ x2:x3\ +\ x1:x2:x3}, a model that contains main effects as well as two-way and three-way interaction effects. The \texttt{1} can be included in the bracket as \texttt{y \midtilde\ }\texttt{(1\ +\ x1\ +\ x2\ +\ x3)\^{}3} to yield the same result. Perhaps surprisingly, \texttt{y \midtilde\ }\texttt{(0\ +\ x1\ +\ x2\ +\ x3)\^{}3} does not include the intercept in the fitted model, since \texttt{0} is converted to \texttt{-1} and carried outside the bracket and power operator. The formula simplification rule, say for \texttt{y \midtilde\ }\texttt{(0\ +\ x1\ +\ x2\ +\ x3)\^{}3}, in \texttt{R} can be found by

\begin{knitrout}
	\definecolor{shadecolor}{rgb}{0.969, 0.969, 0.969}\color{fgcolor}\begin{kframe}
		\begin{alltt}
			\hlkwd{formula}\hlstd{(}\hlkwd{terms}\hlstd{(y} \hlopt{\mytilde} \hlstd{(}\hlnum{0} \hlopt{+} \hlstd{x1} \hlopt{+} \hlstd{x2} \hlopt{+} \hlstd{x3)}\hlopt{^}\hlnum{3}\hlstd{,} \hlkwc{simplify} \hlstd{=} \hlnum{TRUE}\hlstd{))}
		\end{alltt}
		\begin{verbatim}
		## y ~ x1 + x2 + x3 + x1:x2 + x1:x3 + x2:x3 + x1:x2:x3 - 1
		\end{verbatim}
	\end{kframe}
\end{knitrout}

\hypertarget{trees}{%
	\subsection{Trees Volume: Linear Model}\label{trees}}

The \texttt{trees} data set \citep[original data source from][built-in data in R]{minitab} contain 31 observations with 3 numerical variables. The model shown in Figure \ref{fig:symbolic-lm} is a linear model in \eqref{eq:lmm} with the \(31\times 3\) design matrix \(\mathbf{X} = \begin{bmatrix}\boldsymbol{1}_{31} & \log(\boldsymbol{x}_1) & \log(\boldsymbol{x}_2)\end{bmatrix}\), where \(\boldsymbol{x}_1\) is the tree height and \(\boldsymbol{x}_2\) is the tree diameter (named \texttt{Girth} in the data). Finally, \(\boldsymbol{y}\) is the log of the volume of the tree. 

In Figure~\ref{fig:symbolic-lm}, the connection of the data column names to symbolic model formula and its resemblance to the model equation is immediately obvious. As discussed before, transformations may be saved for later analysis using the symbolic model formulae (e.g.~prediction in original scale), however, this likely requires manual recovery when the API requires design matrix as input. 

\hypertarget{grass}{%
	\subsection{Herbicide: Categorical Variable}\label{grass}}

The \texttt{herbicide} data set \citep[original source from R. Hull, Rothamsted Research, data sourced from][]{Welham2015} contains 135 observations with 1 numerical variable (weight response) and 3 categorical variables: block, herbicide, and population of black-grass with 5, 3 and 9 levels respectively. The experiment employed has a factorial treatment structure (i.e.~27 treatments which are combinations of herbicide and population), with the complete set of treatment randomised within each of the five blocks (i.e.~it employs a randomised complete block design).

The model in Figure \ref{fig:symbolic-lm-factor} is a linear model to the square root of the weight of the black-grass with the design matrix \(\mathbf{X} = \begin{bmatrix}\boldsymbol{1}_{135} & \boldsymbol{x}_1 & \cdots & \boldsymbol{x}_{30}\end{bmatrix}\), where \(\boldsymbol{x}_1, ..., \boldsymbol{x}_4\) are dummy variables for \texttt{Block} \texttt{B2}, \texttt{B3} and \texttt{B4}, $\boldsymbol{x}_5, ..., \boldsymbol{x}_{12}$ are dummy variables for \texttt{Population} \texttt{P2} to \texttt{P9}, $\boldsymbol{x}_{13}$ and $\boldsymbol{x}_{14}$ are dummy variables for \texttt{Herbicide} \texttt{B} and \texttt{C}, and $\boldsymbol{x}_{15}, ..., \boldsymbol{x}_{30}$ are dummy variables for the corresponding interaction effects. Alternatively, the model can be written via the  ANOVA model specification,
\[y_{ijk} = \mu + \gamma_k + \alpha_i + \beta_j + (\alpha\beta)_{ij} + e_{ijk}, \]
where index \(i\) denotes for level of population, index \(j\) for level of herbicide and index \(k\) for the replicate block. With dummy variables, the relevant constraints are \(\alpha_1 = \beta_1 = \gamma_1 = (\alpha\beta)_{1j}=(\alpha\beta)_{i1} = 0\). This form is equivalent to the linear regression model given in equation~\eqref{eq:lm} with the fixed effects vector $$\boldsymbol{\beta} = (\mu, \gamma_2,... , \gamma_5, \alpha_2, \alpha_3, ... , \alpha_9, \beta_2, \beta_3, (\alpha\beta)_{22}, (\alpha\beta)_{23}, ..., (\alpha\beta)_{93})^\top.$$

\hypertarget{intercept}{%
	\subsection{Specification of intercept}\label{intercept}}

\citet{Wilkinson1973} described many of the operators and evaluation rules associated with symbolic model formulae, that to this day remain a mainstay of \texttt{R} as well as other languages. These include simplification rules such as \texttt{y} \texttt{\midtilde\ }\texttt{x\ +\ x} and \texttt{y} \texttt{\midtilde\ }\texttt{x:x} to \texttt{y} \texttt{\midtilde\ }\texttt{x}. Their description however did not include any discussion about the intercept. The symbolic evaluation rules governing the intercept are classified as special cases in the current implementation of \texttt{R}, although they may not be as overly intuitive on first glance, e.g.

\begin{itemize}
	\tightlist
	\item
	\texttt{y \midtilde 1:x} simplifies to \texttt{y} \texttt{\midtilde\ }\texttt{1}, although one may expect \texttt{y} \texttt{\midtilde\ }\texttt{x};
	\item
	\texttt{y} \texttt{\midtilde\ }\texttt{1*x} simplifies to \texttt{y} \texttt{\midtilde\ }\texttt{1}, which may be surprising in light of the proceeding point;
	\item
	\texttt{y} \texttt{\midtilde\ }\texttt{x*1} simplifies to \texttt{y} \texttt{\midtilde\ }\texttt{x}, which makes the cross operator unsymmetric for this special case.
\end{itemize}

Further ambiguity arises when we consider cases where we wish to explicitly remove the intercept, e.g.

\begin{itemize}
	\tightlist
	\item
	\texttt{y} \texttt{\midtilde\ }\texttt{-1:x} simplifies to the nonsensical \texttt{y} \texttt{\midtilde\ }\texttt{1\ -\ 1}, which is equivalent to \texttt{y} \texttt{\midtilde\ }\texttt{0},
	\item
	\texttt{y} \texttt{\midtilde\ }\texttt{1\ +\ (-1\ +\ x)} simplifies to \texttt{y} \texttt{\midtilde\ }\texttt{x\ -\ 1}.
\end{itemize}

The last point was raised by \citet{patsy}, and subsequently the formula evaluation differs in the \texttt{patsy} \texttt{Python} library on this particular aspect. These complications arise due to the explicit specification for removing the intercept. Furthermore, the symbolic model formulae that includes \texttt{-1} or \texttt{0} removes the resemblance to the model equation, detracting from the aim of symbolic model formula to make model formulation straightforward and accessible for practitioners. It should be noted, however, that these cases are all somewhat contrived and would rarely be used in practice.

\hypertarget{lmm}{%
	\section{Linear Mixed Models}\label{lmm}}

Consider a \(n\)-vector of response \(\boldsymbol{y}\), which is modelled as
\begin{equation}
\boldsymbol{y} = \mathbf{X}\boldsymbol{\beta} + \mathbf{Z}\boldsymbol{b} + \boldsymbol{e},
\label{eq:lmm}
\end{equation}
where the \(\mathbf{X}\) is the design matrix for the fixed effects coefficients \(\boldsymbol{\beta}\); \(\mathbf{Z}\) is the design matrix of the random effects coefficients \(\boldsymbol{b}\), and \(\boldsymbol{e}\) is the vector of random errors. We typically assume that the random effects and errors are independent of each other and both multivariate normally distributed,
\[\begin{bmatrix}\boldsymbol{b}\\\boldsymbol{e}\end{bmatrix}\sim N\left(\begin{bmatrix}\boldsymbol{0}\\\boldsymbol{0}\end{bmatrix}, \begin{bmatrix}\mathbf{G} & \mathbf{0} \\ \mathbf{0} & \mathbf{R} \end{bmatrix}\right)\]
where \(\mathbf{G}\) and \(\mathbf{R}\) are the covariance matrices of \(\boldsymbol{b}\) and \(\boldsymbol{e}\), respectively. 

In Section~\ref{examples}, we present examples with different variables and structures for model \eqref{eq:lmm}. In the next sections, we briefly describe and contrast the fitting functions \texttt{lmer} and \texttt{asreml} from the \texttt{lme4},  \texttt{asreml} \texttt{R}-packages, respectively.

\hypertarget{lme4::lmer}{%
	\subsection{\texorpdfstring{\texttt{lme4}}{lme4}}\label{lme4}}

The \texttt{lme4} \texttt{R} package fits a LMM with the function \texttt{lmer}. The API consists of a \emph{single} formula that extends the linear model formula as follows -- the random effects are added by surrounding the term in round brackets with grouping structure specified on the right of the vertical bar, and the random terms within each group on the left of the vertical bar, e.g. \texttt{(formula\ \textbar{}\ group)}. The \texttt{formula} is evaluated under the same mechanism for symbolic model formula as linear models in Section \ref{lm}, with \texttt{group} specific effects from \texttt{formula}. These \texttt{group} specific effects are assumed to be normally distributed with zero mean and unstructured variance, as given above in \eqref{eq:lmm}. Examples of its use are provided in Section \ref{examples}.

\hypertarget{asreml::asreml}{%
	\subsection{\texorpdfstring{\texttt{asreml}}{asreml}}\label{asreml}}

In \texttt{asreml}, the random effects are specified as another formula to the argument \texttt{random}. One of the main strength of LMM specification in \texttt{asreml}, in contrast to \texttt{lme4} in wide array of flexible covariance structures. The full list of covariance structures available in \texttt{asreml} Version 3 are given in \citet{Butler2009}; \texttt{asreml} version 4 has some slight differences as outlined in \citet{Butler2018}, although the main concept is similar: variance structures are specified with function-like terms in the model formulae, e.g. \texttt{us(factor)} will fit a \texttt{factor} effect with unstructured covariance matrix; \texttt{diag(factor)} will fit a \texttt{factor} effect with diagonal covariance matrix, i.e.~zero off-diagonal and different parameterisation in the diagonal elements. Note \texttt{factor} corresponds to a categorical variable in the data; see Section \ref{examples} for examples of its usage.

\hypertarget{examples}{%
	\section{Motivating Examples for LMMs}\label{examples}}

This section presents motivating examples with model specification by \texttt{lmer} or \texttt{asreml}. It should be noted that the models are not advocated to be ``correct'', but rather a plausible model that a practitioner may consider in light of the data and design. For succinctness, we omit all \texttt{data} argument to model fit functions. Also, this paper uses \texttt{lme4} version 1.1.21; \texttt{pedigreemm} version 0.3.3 and \texttt{asreml} version 3. 

\hypertarget{chick}{%
	\subsection{Chicken Weight: Longitudinal Analysis}\label{chick}}

The chicken weight data is originally sourced from \citet{chickendata} and found as a built-in data set in \texttt{R}. It consists of the weights of 50 chickens tracked over regular time intervals (not all weights at each time points are observed). Each chicken are fed one of 4 possible diets. 

In this experiment, we are interested in the influence of different diets on chicken weight. We can model the weight of each chicken over time that includes diet effect, overall intercept and slope for time. Fitting these effects as fixed and assuming that the error is i.i.d. means that the observations from same chicken are uncorrelated and there is no variation for the intercept and slope between chickens. This motivates the inclusion of random intercept and random slope for each chicken. More explicitly, and using an ANOVA model specification, the weight may be modelled as
\begin{equation}
y_{ij} = \beta_0 + \beta_1x_{ij} + \alpha_{T(i)} + b_{0i} + b_{1i}x_{ij} + e_{ij}, \label{eq:chickmodel}
\end{equation}
where \(y_{ij}\) is the weight of the \(i\)-th chicken at time index \(j\), \(x_{ij}\) is the days since birth at time index \(j\) for the \(i\)-th chicken, \(b_{0i}\) and \(b_{1i}\) are the random intercept and random time slope effects for the \(i\)-th chicken, \(\beta_0\) and \(\beta_1\) are the overall fixed intercept and fixed time slope, and $e_{ij}$ is the random error.

The above model is incomplete without distributional assumptions for the random effects. As intercept and slope clearly measure different units, the variance will be on different scales. Furthermore, we make an assumption that the random intercept and random slope are correlated within the same chicken, but independent across chickens. With the typical assumption of mutual independence of random effects and random error, and normally and identically distributed (NID) effects, we thus have the distribution assumptions,
\begin{equation}
\begin{bmatrix}b_{0i} \\ b_{1i}\end{bmatrix}\sim NID\left(\begin{bmatrix}0\\0 \end{bmatrix}, \begin{bmatrix}\sigma_0^2 & \sigma_{01} \\ \sigma_{01} & \sigma^2_1\end{bmatrix}\right)\qquad\text{and}\qquad e_{ij}\sim NID(0, \sigma^2).\label{eq:NIDassum}
\end{equation}

If the effects in model \eqref{eq:chickmodel} are vectorised as in model \eqref{eq:lmm} with $$\boldsymbol{b} = (b_{01}, b_{02}, ..., b_{0,50}, b_{11}, b_{12}, ..., b_{1,50})^\top\quad\text{and}\quad \boldsymbol{e} = (e_{ij}),$$
then the model assumption \eqref{eq:NIDassum} can also be written as 
$$\boldsymbol{b}  \sim N\left(\boldsymbol{0}, \begin{bmatrix}\sigma_0^2 & \sigma_{01} \\ \sigma_{01} & \sigma^2_1\end{bmatrix}\otimes \mathbf{I}_{m} \right)\quad\text{and}\quad \boldsymbol{e}\sim N(\boldsymbol{0}, \sigma^2\mathbf{I}_n)$$
where $\otimes$ is the Kronecker product. In \texttt{asreml}, a separable covariance structure, $\mathbf{\Sigma}_1 \otimes \mathbf{\Sigma}_2$, is specified by the use of an interaction operator where the dimensions and structures of $\mathbf{\Sigma}_1$ and $\mathbf{\Sigma}_2$ are specified by the \texttt{factor} input or its number of levels and the function that wraps the \texttt{factor}, e.g. \texttt{us(2):id(50)} is equivalent to $\mathbf{\Sigma}_{2\times 2}\otimes \mathbf{I}_{50}$ where $\mathbf{\Sigma}_{2\times 2}$ is a $2\times 2$ unstructured covariance matrix.

\begin{figure}
	\includegraphics[width=0.9\linewidth,fbox]{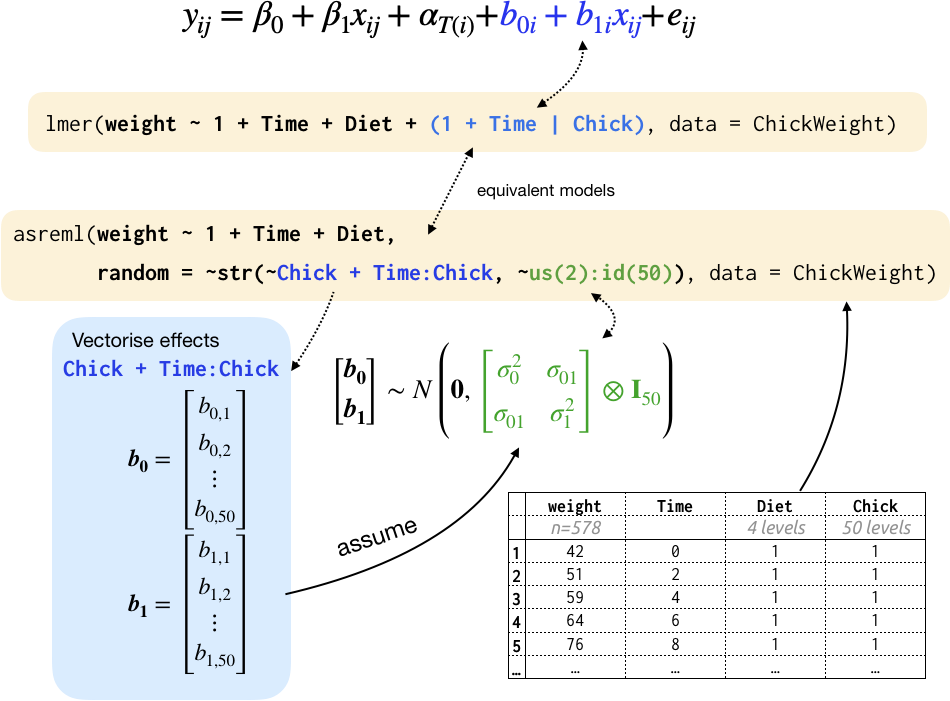} \caption{This figure shows a longitudinal analysis of the chicken data (see Section \ref{chick}). The index form of the model equation shows direct resemblance for symbolic model formula in \texttt{lmer} for the fixed and random effects, however, its covariance form is not as easily inferred. In contrast, the symbolic model formula in \texttt{asreml} show resemblance of the covariance structure specified in the second argument of \texttt{\mytilde str}, however, the corresponding random effects specified in the first argument of \texttt{\mytilde str} must be vectorised as show in the above figure and requires implicit knowledge of the Kronecker product of relevant matrices. }\label{fig:symbolic-lmm}
\end{figure}

The symbolic model formulae that encompasses the model \eqref{eq:chickmodel} coupled with assumption in \eqref{eq:NIDassum} for \texttt{lmer} and \texttt{asreml} are shown in Figure~\ref{fig:symbolic-lmm}. The two symbolic model formulae share the same syntax for fixed effects, however, in this case the random effects syntax is more verbose for \texttt{asreml}.

One may wish to modify their assumption such that now we assume

$$\boldsymbol{b}  \sim N\left(\boldsymbol{0}, \begin{bmatrix}\sigma_0^2 & 0 \\ 0 & \sigma^2_1\end{bmatrix}\otimes \mathbf{I}_{m} \right),$$

That is, the random slope and random intercept are assumed to be uncorrelated. This uncorrelated model may be specified in \texttt{lme4} by replacing \texttt{|} with \texttt{||} as below.
\begin{knitrout}
	\definecolor{shadecolor}{rgb}{0.969, 0.969, 0.969}\color{fgcolor}\begin{kframe}
		\begin{alltt}
			\hlkwd{lmer}\hlstd{(weight} \hlopt{\mytilde} \hlnum{1} \hlopt{+} \hlstd{Time} \hlopt{+} \hlstd{Diet} \hlopt{+} \hlstd{(}\hlnum{1} \hlopt{+} \hlstd{Time} \hlopt{||} \hlstd{Chick))}
		\end{alltt}
	\end{kframe}
\end{knitrout}
It should be noted that the effects specified on the LHS of the \texttt{||} are uncorrelated if the variables are numerical only; we refer to the example in Table~\ref{tab:gxe} for a case where this does not work when the variable is a factor.

\begin{knitrout}
	\definecolor{shadecolor}{rgb}{0.969, 0.969, 0.969}\color{fgcolor}\begin{kframe}
		\begin{alltt}
			\hlkwd{lmer}\hlstd{(weight} \hlopt{\mytilde} \hlnum{1} \hlopt{+} \hlstd{Time} \hlopt{+} \hlstd{Diet} \hlopt{+} \hlstd{(}\hlnum{1} \hlopt{|} \hlstd{Chick)} \hlopt{+} \hlstd{(}\hlnum{0} \hlopt{+} \hlstd{Time} \hlopt{|} \hlstd{Chick))}
		\end{alltt}
	\end{kframe}
\end{knitrout}

The same model is specified as below for \texttt{asreml} where now \texttt{us(2)} is replaced with \texttt{diag(2)}. The correspondence to the covariance structure is more explicit, but again involves the random effects being (implicitly) vectorised as show in Figure~\ref{fig:symbolic-lmm} and care is needed with orders of separable structure.
\begin{knitrout}
	\definecolor{shadecolor}{rgb}{0.969, 0.969, 0.969}\color{fgcolor}\begin{kframe}
		\begin{alltt}
			\hlkwd{asreml}\hlstd{(weight} \hlopt{\mytilde} \hlnum{1} \hlopt{+} \hlstd{Time} \hlopt{+} \hlstd{Diet,}
			\hlkwc{random}\hlstd{=}\hlopt{\mytilde} \hlkwd{str}\hlstd{(}\hlopt{\mytilde}\hlstd{Chick} \hlopt{+} \hlstd{Chick}\hlopt{:}\hlstd{Time,} \hlopt{\mytilde}\hlkwd{diag}\hlstd{(}\hlnum{2}\hlstd{)}\hlopt{:}\hlkwd{id}\hlstd{(}\hlnum{50}\hlstd{)))}
		\end{alltt}
	\end{kframe}
\end{knitrout}

\hypertarget{atrial}{%
	\subsection{Field Trial: Covariance Structure}\label{atrial}}

In this example, we consider wheat yield data sourced from the \texttt{agridat} \texttt{R}-package \citep{agridat}, which  originally appeared in \citet{Gilmour1997}. This data set consists of $n=330$ observations from a near randomised complete block experiment with $m=107$ varieties, of which 3 varieties have 6 replicates while the rest have 3 replicates. The field trial that the yield data was collected from was laid out in a rectangular array with $r=22$ rows and $c=15$ columns. Each of the variety replicates are spread uniformly to $b=3$ blocks. The columns 1-5, columns 6-10 and columns 11-15 form three equal blocks of contiguous area within the field trial. The data frame \texttt{gilmour.serpentine} contains the columns for \texttt{yield}, \texttt{gen} (variety), \texttt{rep} (block), \texttt{col} (column) and \texttt{row}. Further columns \texttt{colf} and \texttt{rowf}, which are factor versions of \texttt{col} and \texttt{row}, have also been added.

We may model the yield observations $\boldsymbol{y}$, ordered by the rows within columns, using the model \eqref{eq:lmm} where here $\boldsymbol{\beta}$ is the $b$-vector of replicate block effects and $\boldsymbol{b}$ is the $m$-vector of variety random effects. We consider next a few potential covariance structures for $\boldsymbol{b}$ and $\boldsymbol{e}$.

\subsubsection{Scaled identity structure.}

One of the simplest assumptions to make would be to assume that $var(\boldsymbol{b}) = \mathbf{G} = \sigma_g^2 \mathbf{I}_m$ i.e., a scaled identity structure. We may additionally assume that $var(\boldsymbol{e}) = \sigma^2\mathbf{I}_n$. In \texttt{lmer}, this is fitted as below. 

\begin{knitrout}
	\definecolor{shadecolor}{rgb}{0.969, 0.969, 0.969}\color{fgcolor}\begin{kframe}
		\begin{alltt}
			\hlkwd{lmer}\hlstd{(yield} \hlopt{\mytilde} \hlnum{0} \hlopt{+} \hlstd{rep} \hlopt{+} \hlstd{(}\hlnum{1} \hlopt{|} \hlstd{gen))}
		\end{alltt}
	\end{kframe}
\end{knitrout}

To elaborate further, \texttt{lmer} specifies a random intercept for each variety. This variety intercept will each be assumed to arise from $NID(0, \mathbf{\Sigma}_{1\times 1})$ where $\mathbf{\Sigma}_{1\times 1}$ is a $1\times 1$ unstructured variance matrix (essentially a single parameter variance component).

The same model is fitted in \texttt{asreml} as below. Particularly, \texttt{idv(gen)} signifies a vector of variety effects with \texttt{idv} variance structure, i.e. a scaled identity structure. This is the default structure in \texttt{asreml}, and so omitting variance structure, \texttt{random = \mytilde gen}, results in the same fit.

\begin{knitrout}
	\definecolor{shadecolor}{rgb}{0.969, 0.969, 0.969}\color{fgcolor}\begin{kframe}
		\begin{alltt}
			\hlkwd{asreml}\hlstd{(yield} \hlopt{\mytilde} \hlnum{0} \hlopt{+} \hlstd{rep,} \hlkwc{random} \hlstd{=} \hlopt{\mytilde}\hlkwd{idv}\hlstd{(gen))}
		\end{alltt}
	\end{kframe}
\end{knitrout}

\subsubsection{Crossed random effects.} 

Field trials often employ rows and/or columns as blocking factors in the experimental design. Furthermore, it is common practice that the management practices of field experiments follow some systematic routine, e.g., harvesting may occur in a serpentine fashion from the first to the last row. These occasionally introduce obvious unwanted noise in the data that are often removed by including random row or column effects assuming that they are i.i.d. for simplicity. These so-called crossed random effects are fitted as below for \texttt{lmer} and \texttt{asreml}.

\begin{knitrout}
	\definecolor{shadecolor}{rgb}{0.969, 0.969, 0.969}\color{fgcolor}\begin{kframe}
		\begin{alltt}
			\hlkwd{lmer}\hlstd{(yield} \hlopt{\mytilde} \hlnum{0} \hlopt{+} \hlstd{rep} \hlopt{+} \hlstd{(}\hlnum{1} \hlopt{|} \hlstd{gen)} \hlopt{+} \hlstd{(}\hlnum{1} \hlopt{|} \hlstd{rowf)} \hlopt{+} \hlstd{(}\hlnum{1} \hlopt{|} \hlstd{colf))}
		\end{alltt}
	\end{kframe}
\end{knitrout}

\begin{knitrout}
	\definecolor{shadecolor}{rgb}{0.969, 0.969, 0.969}\color{fgcolor}\begin{kframe}
		\begin{alltt}
			\hlkwd{asreml}\hlstd{(yield} \hlopt{\mytilde} \hlnum{0} \hlopt{+} \hlstd{rep,} \hlkwc{random} \hlstd{=} \hlopt{\mytilde}\hlkwd{idv}\hlstd{(gen)} \hlopt{+} \hlkwd{idv}\hlstd{(rowf)} \hlopt{+} \hlkwd{idv}\hlstd{(colf))}
		\end{alltt}
	\end{kframe}
\end{knitrout}

\subsubsection{Error covariance structure.}

A field trial is often laid out in a rectangular array and observations from each plot indexed by row and column within this array. Consequently, the assumption that $var(\boldsymbol{e}) = \sigma^2\mathbf{I}_n$ may be restrictive when there is likely to be some sort of spatial correlation, i.e. plots that are geographically closer would be similar than plots further apart. A range of models may be considered for this potential correlation. In practice, a separable autoregressive process of order one, denoted AR1$\times$AR1, has worked well as a compromise between parsimony and flexibility as a structure \citep{Gilmour1997}. More specifically, we assume $var(\boldsymbol{e}) = \sigma^2\mathbf{\Sigma}_c\otimes \mathbf{\Sigma}_r$ where $\mathbf{\Sigma}_c$ is a $c \times c$ matrix with $(i,j)$-th entry of $\mathbf{\Sigma}_c$ given as $\rho_c^{|i - j|}$ with autocorrelation parameter $\rho_c$, and a similar definition holds for $r\times r$ matrix $\mathbf{\Sigma}_r$ except the autocorrelation parameter is denoted bvy $\rho_r$. This model is fitted in \texttt{asreml} by supplying a symbolic formula, \texttt{~ar1(colf):ar1(rowf)}, to the argument \texttt{rcov} as below.

\begin{knitrout}
	\definecolor{shadecolor}{rgb}{0.969, 0.969, 0.969}\color{fgcolor}\begin{kframe}
		\begin{alltt}
			\hlkwd{asreml}\hlstd{(yield} \hlopt{\mytilde} \hlnum{0} \hlopt{+} \hlstd{rep,} \hlkwc{random} \hlstd{=} \hlopt{\mytilde}\hlstd{gen,} \hlkwc{rcov} \hlstd{=} \hlopt{\mytilde}\hlkwd{ar1}\hlstd{(colf)}\hlopt{:}\hlkwd{ar1}\hlstd{(rowf))}
		\end{alltt}
	\end{kframe}
\end{knitrout}

Here, the \texttt{ar1} specifies an autoregressive process of order one with dimension given by number of levels in \texttt{rowf} and \texttt{colf}. It is important to note that \texttt{ar1} denotes a correlation matrix and a covariance matrix may be specified by \texttt{ar1v}. Care needs to be taken in covariance specification for separable models, as clearly there is a lack of variance parameter(s) where $\mathbf{\Sigma}_1$ and $\mathbf{\Sigma}_2$ are both correlation structures only, while if both are covariance structure then the model is over-parameterised and unidentifiable. In the error structure of \texttt{rcov}, this is taken care of such that 
\texttt{rcov = \mytilde ar1v(colf):ar1(rowf)}, \texttt{rcov = \mytilde ar1(colf):ar1v(rowf)} and \texttt{rcov = \mytilde ar1(colf):ar1(rowf)} will fit all the same model. It should be noted that this is not the case for separable covariance structures specified in \texttt{random} effects.

In comparison, the more restrictive API of \texttt{lmer} function does not allow the assumption on the random effects to be relaxed from $var(\boldsymbol{e}) = \sigma^2\mathbf{I}_n$. One may of course introduce a random effect, $\boldsymbol{b}_e \sim N(\boldsymbol{0}, \sigma^2\mathbf{\Sigma}_c\otimes \mathbf{\Sigma}_r)$, and assume $\boldsymbol{e} \sim N(\boldsymbol{0}, \sigma^2\mathbf{I}_n)$. However, this separable covariance structure also can not be specified within \texttt{lmer} function.


\subsubsection{Known covariance structure.} 

Often in plant breeding trials, the varieties of interest have some shared ancestry. This is captured in the form of pedigree data that contains 3 columns: individual ID, mother's ID and father's ID. The related structure is commonly captured by the use of a numerator relationship matrix, denoted here has $\mathbf{A}$ \citep{Mrode2014}. For example, suppose that individuals $i$ and $j$ are full-siblings. Then the corresponding $(i,j)$-th entry in $\mathbf{A}$ is 0.5 (i.e., the average probability that a randomly drawn allele from individual $i$ is identical by descent to the randomly drawn allele at the same autosomal locus from individual $j$). 

With the additional information above, we may assume that $var(\boldsymbol{b}) = \sigma^2_g \mathbf{A}$ to exploit this \textit{known} relatedness structure between varieties. The symbolic model formulae in \texttt{lme4} alone is unable to specify this model and, an extension \texttt{R} package \texttt{pedigreemm} \citep{pedigreemm-paper} is required. The pedigree data is parsed to make an object of \texttt{pedigree} class, which we refer to here as \texttt{ped}. This object \texttt{ped} is then included as part of the input in the main fitting function \texttt{pedigreemm}, as depicted below.

\begin{knitrout}
	\definecolor{shadecolor}{rgb}{0.969, 0.969, 0.969}\color{fgcolor}\begin{kframe}
		\begin{alltt}
			\hlkwd{pedigreemm}\hlstd{(yield} \hlopt{\mytilde} \hlnum{0} \hlopt{+} \hlstd{rep} \hlopt{+} \hlstd{(}\hlnum{1} \hlopt{|} \hlstd{gen),}
			\hlkwc{pedigree} \hlstd{=} \hlkwd{list}\hlstd{(}\hlkwc{gen} \hlstd{= ped))}
		\end{alltt}
	\end{kframe}
\end{knitrout}

In \texttt{asreml}, the fit is similar to the above, but the factor with the known covariance structure must be wrapped in \texttt{giv} with argument \texttt{ginverse} providing a named list with the inverse of the $\mathbf{A}$ in a sparse format, i.e. a data frame of 3 columns that consists the row and the column index of $\mathbf{A}$ and its corresponding value in $\mathbf{A}$ provided that the value is non-zero.

\begin{knitrout}
	\definecolor{shadecolor}{rgb}{0.969, 0.969, 0.969}\color{fgcolor}\begin{kframe}
		\begin{alltt}
			\hlkwd{asreml}\hlstd{(yield} \hlopt{\mytilde} \hlnum{0} \hlopt{+} \hlstd{rep,} \hlkwc{random} \hlstd{=} \hlopt{\mytilde}\hlkwd{giv}\hlstd{(gen),}
			\hlkwc{ginverse} \hlstd{=} \hlkwd{list}\hlstd{(}\hlkwc{gen} \hlstd{= Ainv))}
		\end{alltt}
	\end{kframe}
\end{knitrout}

\hypertarget{MET}{%
	\subsection{Multi-Environmental Trial: Separable Structure}\label{MET}}

In the final example, we consider CIMMYT Australia ICARDA Germplasm Evaluation (CAIGE) bread wheat yield 2016 data \citep{CAIGE2016}, which consists of $t=7$ sites across Australia, where the overall aim is to select the best genotype (\texttt{gen}). There were $m=240$ genotype tested across seven trials and 252-391 plots, with a total of $n=2127$ yield observations. Each trial employed a partially replicated ($p$-rep) design \citep{Cullis2006}, with $p$ ranging from 0.23 to 0.39.

Fitting a model to a model should take into account the differential mean yield across sites, and allow for different genotypic variations by site. For simplicity, we ignore other variations for now. In turn, the LMM formulation in equation \eqref{eq:lmm} may be used where $\boldsymbol{y}$ is the vector of yield (ordered rows within columns within sites); $\boldsymbol{\beta}$ is the $t$-vector of site effects; and $\boldsymbol{b}$ is the $mt$-vector of genotype-by-site effects (ordered by genotype within site). There are a number of distributions that may be considered for $\boldsymbol{b}$, as explained below.

We may consider a separable model such that $\boldsymbol{b} \sim N(\boldsymbol{0}, \mathbf{\Sigma}_s \otimes \mathbf{\Sigma}_g)$, where $\mathbf{\Sigma}_s$ and $\mathbf{\Sigma}_g$ are a $t\times t$ and $m\times m$ matrices, respectively. We may further assume that $\mathbf{\Sigma}_g$ has a known structure similar to Section~\ref{atrial}, but for simple illustration here we will assume that the genotypes are independent, i.e. $\mathbf{\Sigma}_g = \mathbf{I}_m$. Also, we may assume that $\mathbf{\Sigma}_s = \text{diag}\left(\sigma^2_{g1}, \sigma^2_{g2}, \cdots, \sigma^2_{gt}\right)$, i.e. a diagonal matrix with different variance paramterisation for each site, thus allowing for different genotypic variance at each site. This model can be fitted as below in \texttt{asreml}.

\begin{knitrout}
	\definecolor{shadecolor}{rgb}{0.969, 0.969, 0.969}\color{fgcolor}\begin{kframe}
		\begin{alltt}
			\hlkwd{asreml}\hlstd{(yield} \hlopt{\mytilde} \hlnum{0} \hlopt{+} \hlstd{site,} \hlkwc{random} \hlstd{=} \hlopt{\mytilde} \hlkwd{diag}\hlstd{(site)}\hlopt{:}\hlkwd{id}\hlstd{(gen))}
		\end{alltt}
	\end{kframe}
\end{knitrout}

The same model in \texttt{lmer} is somewhat more involved as shown in Table~\ref{tab:gxe}. 

\begin{table}[]
	\centering
	
	\caption{The table lists the equivalent symbolic model formula in \texttt{lmer} and \texttt{asreml} for the site-by-genotype random effect, $\boldsymbol{b}$ and the corresponding mathematical form of the variance structure of $\boldsymbol{b}$. Here, $\mathbf{\Sigma}_{t\times t}$ is a $t\times t$ unstructured covariance matrix; $\mathbf{D}_{t\times t} = \text{diag}(\sigma^2_{g1}, ..., \sigma^2_{g7})$, a $t\times t$ diagonal covariance matrix; $m$ is the number of genotypes; $t$ is the number of sites; and \texttt{S1} is a $n$-vector where the entry is one if the corresponding observation belongs to site 1 and zero otherwise (similar definitions hold for \texttt{S2}, \ldots, \texttt{S7}). The conversion of the factor \texttt{site} to numerical variables \texttt{S1}, ..., \texttt{S7} is required to have uncorrelated random effects in \texttt{lmer} via the \texttt{||} operator, as per the last row in the table. The \texttt{||} group separation in \texttt{lmer} is only effective when variables on LHS are numerical.}
	\label{tab:gxe}
	\begin{tabular}{l@{\hskip 0.5cm}ll}
		\toprule
		\texttt{lmer} & \texttt{asreml} & $var(\boldsymbol{b})$  \\\toprule
		& \texttt{idv(site):id(geno)}&  \\
		\texttt{(1 | site:geno)} & \texttt{id(site):idv(geno)}&$\sigma^2_g\mathbf{I}_{tm}$  \\
		& \texttt{site:geno}& \\ \midrule
		\texttt{(0 + site | geno)}&  &   \\
		\texttt{(0 + site || geno)} & \texttt{us(site):id(geno)}& $\mathbf{\Sigma}_{t\times t}\otimes \mathbf{I}_{t}$ \\
		\texttt{(0 + S1 + S2 + S3 + S4 + S5 + S6 + S7 | geno)} & & \\
		\midrule
		\texttt{(0 + S1 + S2 + S3 + S4 + S5 + S6 + S7 || geno)} &\texttt{diag(site):id(geno)} & $\mathbf{D}_{t\times t}\otimes \mathbf{I}_{t}$ \\
		\bottomrule
	\end{tabular}
\end{table}

\begin{figure}
	\includegraphics[width=0.9\linewidth,fbox]{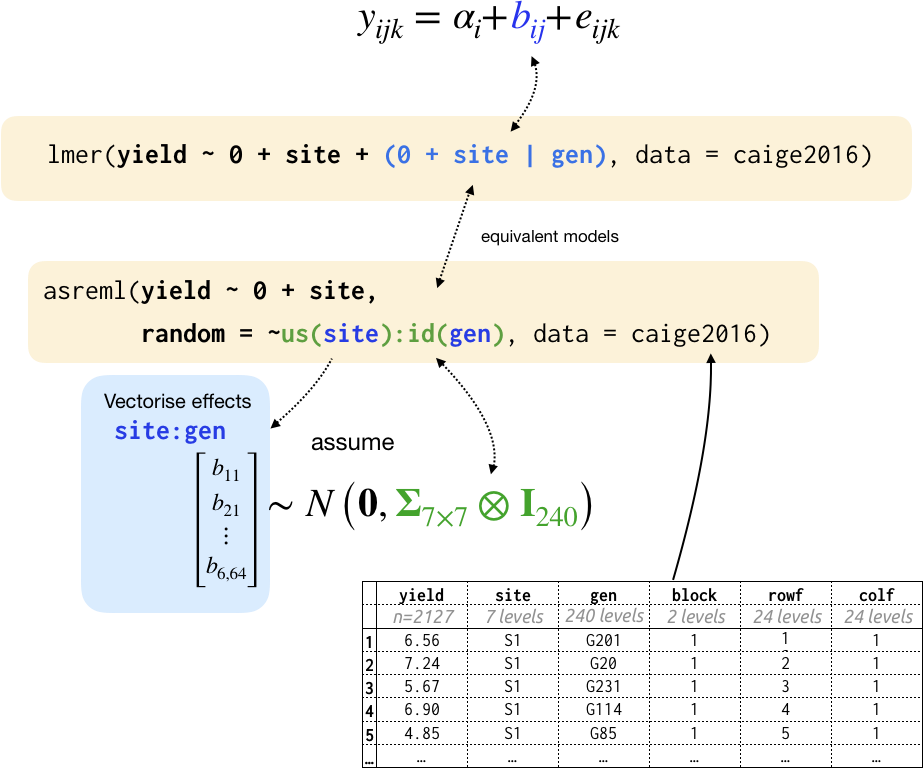} \caption{Depiction of the fit of a simplified LMM for the analysis of the MET data. In modelling the \texttt{site}-by-\texttt{gen} random effect, the variance structure are specified differently using \texttt{lmer} and \texttt{asreml}, where latter shows resemblances of covariance structure written mathematically and when all random effects are vectorised and concatenated, while the former requires some additional computation.}\label{fig:symbolic-lmm2}
\end{figure}

The diagonal model assumes that genotype-by-site effects are uncorrelated across sites for the same genotype. However, a more realistic assumption is to assume that these effects are correlated, thus allowing for different correlation of genotype effects between pair of sites, i.e. we assume that $\mathbf{\Sigma}_s$ is an unstructured covariance matrix. The specification of such model for \texttt{lmer} and \texttt{asreml} is shown in Figure~\ref{fig:symbolic-lmm2}.

A even more realistic model may consider including site-specific random row or column effects, and assuming an AR1$\times$AR1 process for the error covariance at each site as in Section~\ref{atrial}. These are easily included in \texttt{asreml} using the \texttt{at} function within the symbolic model formulae. For example, the inclusion of random row effect at site \texttt{S1} only and AR1$\times$AR1 processes for the error covariance at each site is shown below.

\begin{knitrout}
	\definecolor{shadecolor}{rgb}{0.969, 0.969, 0.969}\color{fgcolor}\begin{kframe}
		\begin{alltt}
			\hlkwd{asreml}\hlstd{(yield} \hlopt{\mytilde} \hlnum{0} \hlopt{+} \hlstd{site,}
			\hlkwc{random} \hlstd{=} \hlopt{\mytilde}\hlkwd{us}\hlstd{(site)}\hlopt{:}\hlkwd{id}\hlstd{(gen)} \hlopt{+} \hlkwd{at}\hlstd{(site,} \hlstr{"S1"}\hlstd{)}\hlopt{:}\hlkwd{idv}\hlstd{(rowf),}
			\hlkwc{rcov} \hlstd{=} \hlopt{\mytilde}\hlkwd{at}\hlstd{(site)}\hlopt{:}\hlkwd{ar1}\hlstd{(colf)}\hlopt{:}\hlkwd{ar1}\hlstd{(rowf))}
		\end{alltt}
	\end{kframe}
\end{knitrout}

The above model cannot be specified using \texttt{lmer}.

\hypertarget{discussion}{%
	\section{Discussion}\label{discussion}}

In fitting statistical models, the user may not necessary understand the full intricacies of model fitting process. However, it is essential that the user understands how to specify the model that they wish to fit and the interpretation from the fit. Symbolic model formulae is a way of bridging the gap between software and mathematical representation of the model, and has been extensively employed in \texttt{R} for this purpose.

In this article, we have extensively compared two widely used LMM \texttt{R}-packages with contrasting model specification in functions: \texttt{lmer} and \texttt{asreml}. Both of these functions use symbolic model formulae to specify the model with \texttt{lmer} taking a more hierarchical approach to random effects specification, while \texttt{asreml} focuses on the covariance structure of the vectorised random effects (and the data for the matter). There are strength and weakness in both approaches as we discuss next.

It is clear from Section~\ref{chick} that a random intercept and random slope model is verbose using the symbolic model formulae of \texttt{asreml}. Specifically, the random effect symbolic formula contains a function \texttt{str} that takes input of two other formula: the first input specifying the random effects, and second input specifying the covariance structure of the vectorised form of random effects specified in the first input. The second input also requires the dimension(s) of covariance structure as input. These number may need manual update when the data is subsequently updated, thus making this symbolic model formula clumsy to use.

On the other hand, the flexibility of \texttt{asreml} is evident in Sections~\ref{atrial} and \ref{MET}, where the LMMs fitted are less easy to pose hierarchically, but the vectorised version of the LMM remains straightforward provided one knows how to establish the set up the structure of the covariance matrices. Put another way, the vast set of in-built pre-defined covariance structures in \texttt{asreml} (e.g. scaled identity, diagonal structure, unstructured, autoregressive process), along with the capacity to modify the error covariance structure and incorporate separable structures makes the model specification embedded in \texttt{asreml} a superior choice here. There are many more pre-defined covariance structures not demonstrated in this paper, and interested readers may refer to \citet{Butler2009, Butler2018}.
By contrast, the lack of flexibility in \texttt{lme4} means that either a more obtuse workaround is required or the precise LMM can not be formulated at all.

That being said, the \citet{brmsjss, brmsr} (\texttt{brms}) make extensive discussion of symbolic model formula and extends on the framework built on \texttt{lme4}. The \texttt{brms} model formulae resembles  \texttt{lme4} and many symbolic model formulae in our examples will be similar. The \texttt{brms} \texttt{R}-package uses a Bayesian approach to fit its models and model specification require further discussion on specifying priors. These discussions are left for future review, although we acknowledge that such extensions may well resolve some of the current limitations of \texttt{lme4} and bridge its gap in flexibility with \texttt{asreml}.

Symbolic model formulae in \texttt{R} is widely used and frameworks to specify mixed models by \texttt{lme4} and \texttt{asreml} (version 3) used for many years. This makes drastic changes difficult for these frameworks. Based on our review, we argue that ideally any new framework for symbolic model formulae should require intercepts to be specified explicitly. As discussed in Section~\ref{intercept}, the default inclusion or explicit removal of intercepts removes the resemblance of symbolic model formulae to the model equation. Currently, the implicit inclusion of intercepts makes certain model formulation unclear and inconsistent across different LMM specifications, e.g. \texttt{(Time | Chick)} in \texttt{lmer} includes random intercept (and slope) for \texttt{Chick}, but the equivalent formulation \texttt{str(\mytilde Chick + Chick:Time, \mytilde diag(2):id(50))} in \texttt{asreml} does not include the random intercept.

There is a trade-off between different types of symbolic model formulae: \texttt{lmer} syntax is no doubt less flexible and may be less intuitive to some, however, with a degree of familiarity pertains as a higher level language for symbolic model formula. For many hierarchical models, the formulation is more elegant and simpler than \texttt{asreml}. However, \texttt{asreml} is more flexible to specify variety of covariance matrices. This strength is predicated on having a deeper understanding of random effects and its covariance structure, and promotes the view of the LMM in a fully vectorised form. 


\section*{Acknowledgement}

This paper benefited from twitter conversation with Thomas Lumley. This paper is made using \texttt{R} Markdown \citep{rmarkdown}. Huge thanks goes to the teams behind \texttt{lme4} and \texttt{asreml} \texttt{R}-packages that make fitting of general LMMs accessible to wider audiences. All materials used to produce this paper and its history of changes can be found on github \url{https://github.com/emitanaka/paper-symlmm}.

\bibliography{biblio}

\end{document}